\documentclass[]{aastex631}

\received{November 1, 2024}
\revised{December 10, 2024}
\accepted{January 20, 2025}
\submitjournal{ApJ}

\shorttitle{Anisotropic model of heliosphere}
\shortauthors{Florinski et al.}

\begin{document}

\title{An anisotropic plasma model of the heliospheric interface}

\correspondingauthor{Vladimir Florinski}
\email{vaf0001@uah.edu}

\author[0000-0001-5485-2872]{Vladimir Florinski}
\affiliation{Department of Space Science, University of Alabama in Huntsville, Huntsville, AL, USA}
\affiliation{Center for Space Plasma and Aeronomic Research, University of Alabama in Huntsville, Huntsville, AL, USA}

\author[0000-0003-3309-1052]{Dinshaw S. Balsara}
\affiliation{Department of Physics and Astronomy, University of Notre Dame, Notre Dame, IN, USA}
\affiliation{ACMS Department, University of Notre Dame, Notre Dame, IN, USA}

\author[0000-0003-2849-9045]{Deepak Bhoriya}
\affiliation{Department of Physics and Astronomy, University of Notre Dame, Notre Dame, IN, USA}

\author[0000-0002-4642-6192]{Gary P. Zank}
\affiliation{Department of Space Science, University of Alabama in Huntsville, Huntsville, AL, USA}
\affiliation{Center for Space Plasma and Aeronomic Research, University of Alabama in Huntsville, Huntsville, AL, USA}

\author[0000-0002-4879-8889]{Shishir Biswas}
\affiliation{Department of Physics and Astronomy, University of Notre Dame, Notre Dame, IN, USA}

\author[0000-0001-8541-7523]{Swati Sharma}
\affiliation{Center for Space Plasma and Aeronomic Research, University of Alabama in Huntsville, Huntsville, AL, USA}

\author[0000-0003-0017-0811]{Sethupathy Subramanian}
\affiliation{Department of Electrical Engineering, Indian Institute of Technology, Dhanbad, India}

\begin{abstract}
We present a pioneering model of the interaction between the solar wind and the surrounding interstellar medium that includes the possibility of different pressures in directions parallel and perpendicular to the magnetic field. The outer heliosheath region is characterized by a low rate of turbulent scattering that would permit development of pressure anisotropy. The effect is best seen on the interstellar side of the heliopause, where a narrow region develops with an excessive perpendicular pressure resembling a plasma depletion layer typical of planetary magnetspheres. The magnitude of this effect for typical heliospheric conditions is relatively small owing to proton-proton collisions. We show, however, that if the circumstellar medium is warm and tenuous, a much broader anisotropic boundary layer can exist, with a dominant perpendicular pressure in the southern hemisphere and a dominant parallel pressure in the north.
\end{abstract}



\section{Introduction} \label{sec:intro}

The heliopause (HP) is the plasma boundary of the solar system, a separatrix layer between the cold, partially ionized and strongly magnetized local interstellar medium (LISM) and the warm inner heliosheath (IHS, the region between the termination shock and the HP). The existence of the heliopause, long since predicted by theory \citep{Parker_1961, Axford_Dessler_Gottlieb_1963} and computer simulations \citep{Baranov_Malama_1993, Pauls_Zank_Williams_1995, Pogorelov_Matsuda_1998}, is now firmly established by in situ observations by NASA’s Voyager 1 and Voyager 2 spacecraft. The two deep space probes encountered a magnetic shear layer characteristic of the heliopause at heliocentric distances of 122 and 119 au, respectively \citep{Burlaga_Ness_Stone_2013, Burlaga_Ness_Berdichevsky_Park_Jian_Szabo_Stone_Richardson_2019}. The outer heliosheath (OHS), a boundary layer between the HP and the bow wave, was found to have a plasma number density of the order of 0.1 cm$^{-3}$, which is about a factor of $\sim 50$ larger than in the inner heliosheath \citep{Gurnett_Kurth_Burlaga_Ness_2013}, and a magnetic field of between 4 and 7 $\mu$G -- two to three times stronger than the field on the solar side of the HP \citep{Burlaga_Ness_Berdichevsky_Jian_Kurth_Park_Rankin_Szabo_2022}.

Voyager 1 does not have a functioning plasma instrument, and the plasma density in the OHS was inferred from electrostatic oscillations thought to be produced by electron beams ahead of propagating shocks \citep{Gurnett_Kurth_2019}, and from weak thermal noise at the local plasma frequency \citep{Burlaga_Kurth_Gurnett_Berdichevsky_Jian_Ness_Park_Szabo_2021}. While the plasma instrument on board Voyager 2 is functional, it is not well suited to perform observations in the OHS because three of its four sensors are pointing in the solar direction \citep{Bridge_Belcher_Butler_Lazarus_Mavretic_Sullivan_Siscoe_Vasyliunas_1977}, where as the plasma flow in the OHS is most probably directed toward the Sun, and because the energy of the protons in the OHS in only a few eV, which is below the lower energy threshold of the instrument. Nonetheless, \citet{Richardson_Belcher_GarciaGalindo_Burlaga_2019} were able to place a constraint on the plasma temperature at 30,000--50,000 K. This is substantially larger than $\sim$7000--9000 K, the temperature in the local interstellar medium (LISM), well upstream of the bow wave, inferred from neutral helium observations \citep{Witte_2004, McComas_Bzowski_Frisch_Fuselier_Kubiak_Kucharek_Leonard_Mobius_Schwadron_Sokol_etal_2015}. Numerical models predict the temperature near the HP to lie in the range between between 15,000 K \citep{Zieger_Opher_Schwadron_McComas_Toth_2013} and 45,000 K \citep{Fraternale_Pogorelov_Bera_2023}.

Voyager 1 measurements beyond the HP showed a general tend of increasing plasma density with heliocentric distance. \citet{Gurnett_Kurth_2019} reported a twofold increase in density, from 0.06 cm$^{-3}$ to 0.12 cm$^{-3}$ over a distance of 24 AU. This observation is reminiscent of plasma depletion layers (PDLs) known to exist between a bow shock and magnetopause of a planet. As first suggested by \citet{Zwan_Wolf_1976}, density depletion is a consequence of magnetic field compression near the nose of the magnetic boundary, squeezing out of the plasma towards the flanks. The opposite trends exhibited by magnetic field strength and density are characteristic of a slow magnetosonic wave. Using a resistive MHD magnetospheric model, \citet{Wang_Raeder_Russell_2004} showed that pressure gradients are primarily responsible for accelerating plasma parcels in the direction parallel to the magnetic field lines. PDLs were subsequently discovered around magnetospheres of Mercury \citep{Gershman_Slavin_Raines_Zurbuchen_Anderson_Korth_Baker_Solomon_2013}, Earth \citep{Phan_Paschmann_Baumjohann_Sckopke_Luhr_1994}, Mars \citep{Oieroset_Mitchell_Phan_Lin_Crider_Acuna_2004}, Jupiter \citep{Tsurutani_Southwood_Smith_Balogh_1993}, Saturn \citep{Masters_Phan_Badman_Hasegawa_Fujimoto_Russell_Coates_Dougherty_2014}, and Neptune \citep{Jasinski_Murphy_Jia_Slavin_2022}.

A distinguishing feature of a PDL is a decrease in parallel pressure $p_\parallel$ compared with perpendicular pressure $p_\perp$ owing to ions with large parallel velocity component escaping the compression region along the field lines leaving the gyrating ions behind. Under such conditions magnetic mirroring becomes the predominant effect pushing the plasma away from the region of strongest magnetic field \citep{Denton_Lyon_1996}. Large values of the pressure anisotropy could lead to mirror or ion cyclotron instability development, and both modes were identified experimentally \citep{Anderson_Fuselier_1993, Violante_BavassanoCattaneo_Moreno_Richardson_1995}.

\citet{Cairns_Fuselier_2017} argued, based on measured Voyager 1 density and magnetic field profiles beyond the HP, that the probe traversed a PDL with a width of 2.6 AU and a depletion factor of 2. Using a global MHD model of the heliosphere, \citet{Pogorelov_Heerikhuisen_Roytershteyn_Burlaga_Gurnett_Kurth_2017} predicted that the plasma density would continue to increase for at least 100 au beyond the HP. This effects, however, is not produced by magnetic forces, but by charge exchange with interstellar neutral atoms. Furthermore, pressure anisotropy is expected to become smaller owing to Coulomb collisions, which are relatively frequent in the OHS, at least compared with the helioosphere. \citet{Fraternale_Pogorelov_2021} estimated the mean free path for proton collisions at $\sim 4$ au and decreasing with heliocentric distance. It is therefore expected that the heliospheric PDL, if it exists, should be relatively narrow, not exceeding a few au in width.

A direct modeling of the global heliospere with an anisotropic pressure tensor has never been attempted. The simplest fluid model that allows for temperature anisotropy is the Chew-Goldberger-Low (CGL) double-adiabatic model \citep{Chew_Goldberger_Low_1956}. Numerous CGL numerical models were developed in the past three decades to study the magnetospheres of Earth and other planets \citep{Erkaev_Farrugia_Biernat_1999, Samsonov_Pudovkin_Gary_Hubert_2001, Hu_Denton_Lin_2010, Meng_Toth_Liemohn_Gombosi_Runov_2012}. Recently, a new CGL model with improved treatment of non-conservative terms and the isotropization process was developed by \citet{Bhoriya_Balsara_Florinski_Kumar_2024}. The purpose of this paper is to apply that model to the heliospheric interface in order to elucidate the possible pressure anisotropies ahead of the HP on the LISM side of the interface. The numerical model presented here draws on our extensive work with geodesic meshes that constitute an efficient framework for modeling of astrophysical systems with a strong degree of central symmetry \citep{Balsara_Florinski_Garain_Subramanian_Gurski_2019, Florinski_Balsara_Garain_Gurski_2020, Subramanian_Balsara_ud_Doula_Gagne_2022}.


\section{Double adiabatic MHD model} \label{sec:model}

The CGL system of equations is obtained from the Boltzmann equation under the assumption of gyrotropy in the electric drift frame via an expansion in $r_g/L\ll 1$, where $r_g$ is the Larmor radius, and $L$ is a characteristic spatial scale of the problem \citep{Chew_Goldberger_Low_1956}. Unlike ideal (isotropic) MHD, the pressure tensor has two distinct components in the directions parallel ($p_\parallel$) and orthogonal ($p_\perp$) to the magnetic field $\mathbf{B}$. In a fixed inertial coordinate frame the pressure tensor has the form
\begin{equation}
\label{eq_ptensor}
\mathbf{p}=p_\perp\mathbf{I}+(p_\parallel-p_\perp)\mathbf{bb},
\end{equation}
where $\mathbf{b}=\mathbf{B}/B$ is the unit vector in the direction of the magnetic field. The resulting system of equations has five physical variables (density $\rho$, velocity $\mathbf{u}$, parallel pressure $p_\parallel$, perpendicular pressure $p_\perp$, and magnetic field $\mathbf{B}$) and consists of nine scalar equations. It cannot be written in a strictly conservative form because of the exchange of the thermal energy $\mathcal{E}$ between the parallel and the perpendicular degrees of freedom. Following \citet{Bhoriya_Balsara_Florinski_Kumar_2024} we retain the MHD conservation laws for mass, momentum, energy (appropriately corrected to account for anisotropic pressure), and magnetic flux, and designate $p_\parallel-p_\perp$ as the extra non-conserved variable. With these assumptions the governing CGL equations are
\begin{equation}
\label{eq_mass_cons}
\frac{\partial\rho}{\partial t}+\nabla\cdot(\rho\mathbf{u})=0,
\end{equation}
\begin{equation}
\label{eq_momentum_conv}
\frac{\partial(\rho\mathbf{u})}{\partial t}+\nabla\cdot\left[\rho\mathbf{uu}+p_\perp\mathbf{I}+(p_\parallel-p_\perp)\mathbf{b}\mathbf{b}-\frac{1}{4\pi}\left(\mathbf{B}\mathbf{B}-\frac{B^2}{2}\textbf{I}\right)\right]=0,
\end{equation}
\begin{equation}
\label{eq_energy_cons}
\frac{\partial\mathcal{E}}{\partial t}+\nabla\cdot\left\lbrace\left(\mathcal{E}+p_\perp+\frac{B^2}{8\pi}\right)\mathbf{u}+\mathbf{u}\cdot\left[(p_\parallel-p_\perp)\mathbf{b}\mathbf{b}-\frac{\mathbf{B} \mathbf{B}}{4\pi}\right]\right\rbrace=0,
\end{equation}
\begin{equation}
\label{eq_deltap}
\frac{\partial(p_\parallel-p_\perp)}{\partial t}+\nabla\cdot\left[(p_\parallel-p_\perp)\mathbf{u}\right]+(2p_\parallel+p_\perp)\mathbf{b}\cdot\nabla\mathbf{u}\cdot\mathbf{b}-p_\perp\nabla\cdot\mathbf{u}=-\frac{p_\parallel-p_\perp}{\tau},
\end{equation}
\begin{equation}
\label{eq_faraday}
\frac{\partial\mathbf{B}}{\partial t}-\nabla\times(\mathbf{u}\times\mathbf{B})=0.
\end{equation}
The energy density in equation (\ref{eq_energy_cons}) is calculated as
\begin{equation}
\mathcal{E}=\frac{\rho u^2}{2}+\frac{p_\parallel}{2}+p_\perp+\frac{B^2}{8\pi}.
\end{equation}
All equations are written in CGS units. The right hand side of equation (\ref{eq_deltap}) contains the relaxation term that models the isotropizing effects of wave-particle interactions and Coulomb collisions on a timescale $\tau$. In what follows we will use parallel and perpendicular plasma betas $\beta_\parallel$ and $\beta_\perp$ as the ratios between the respective proton pressure component and the magnetic pressure $B^2/(8\pi)$, i.e., the electron pressure is considered to be small. We define the pressure anisotropy as $A=p_\perp/p_\parallel$. Readers who would like to learn more about the properties and limitations of the CGL-MHD model are referred to \citet{Hunana_Tenerani_Zank_Khomenko_Goldstein_Webb_Cally_Collados_Velli_Adhikari_2019}.

It is well known that a bi-Maxwellian plasma distribution is subject to several types of instabilities. The CGL theory predicts two instabilities, firehose and mirror, that are associated with the Alfv\'en and the slow magnetosonic waves, respectively \citep{Kato_Tajiri_Taniuti_1966}. In kinetic theory these are replaced by magnetosonic firehose, Alfv\'en firehose, ion cyclotron, and mirror instabilities \citep{Hellinger_Travnicek_Kasper_Lazarus_2006, Bale_Kasper_Howes_Quataert_Salem_Sundkvist_2009}. A fluid model does not need to distinguish between these different modes, but only include the anisotropy thresholds. The following thresholds are commonly used in CGL models:
\begin{equation}
\label{eq_threshold_firehose}
\mathrm{firehose\;instability\;triggered\;when}\quad\beta_\parallel>\beta_\perp+2,
\end{equation}
\begin{equation}
\label{eq_threshold_mirror}
\mathrm{mirror\;instability\;triggered\;when}\quad\beta_\perp>\frac{\beta_\parallel}{2}+\sqrt{\beta_\parallel\left(1+\frac{\beta_\parallel}{4}\right)}.
\end{equation}

The CGL PDE system stops being physically realizable, i.e. hyperbolic and solvable on a computer, if the pressure anisotropy grows without bounds. The model used in this work features an ``elastic fence'' technique introduced earlier in \citet{Bhoriya_Balsara_Florinski_Kumar_2024}. Its design is such that for most of the domain in which the PDE system is physically realizable it allows us to use the physical relaxation time. It is only when the pressure anisotropy very closely approaches the limit that the anisotropy is safely pushed away from going beyond the bounds of physical realizability.
The value of $\tau$ in the model is the timescale for pitch-angle scattering due to ambient turbulence and that of proton-proton collisions. For pitch-angle scattering we may take $\tau=D_{\mu\mu}^{-1}$, where
\begin{equation}
\label{eq_dmumu}
D_{\mu\mu}\simeq\frac{\pi}{2}\frac{\Omega^2}{vB^2}P(k_\mathrm{res}),
\end{equation}
where $v$ is the velocity of a thermal proton, $\Omega$ is the cyclotron frequency, and $P(k_\mathrm{res})$ is the magnetic power spectral density (PSD) at the resonant wavenumber $k_\mathrm{res}\sim\Omega/v$, which follows from the quasi-linear theory of wave-particle interactions \citep[e.g.,][]{Jokipii_1966}. Assuming the PSD is a power law with an index $\gamma$ in the wavenumber range between $l_c^{-1}$ (where $l_c$ corresponds to the outer scale or the correlation length of the turbulence that are typically closely related) and infinity, and that the magnetic variance in the inertial range is equal to $aB^2$, where $a\sim 0.1$--0.2, we may write
\begin{equation}
\label{eq_tau_scat}
\tau_\mathrm{scat}\simeq\frac{4}{\pi(\gamma-1)a\Omega}\left(\frac{l_c}{r_g}\right)^{\gamma-1}.
\end{equation}
This expression has three variables: the magnetic field $B$, the correlation length $l_c$, and the thermal speed $v$ that enters into $r_g$. The value of the correlation length must be assumed. It is often inversely proportional to $B$ because a compression of the field also compresses the fluctuations normal to the field, which are predominant. Suppose then that
\begin{equation}
\label{eq_lcb}
l_c=l_{c0}\frac{B_0}{B},
\end{equation}
where $l_{c0}$ and $B_0$ are some reference values (that are different in different regions). Then, assuming $\gamma=5/3$, we can write equation (\ref{eq_tau_scat}) as
\begin{equation}
\label{eq_tau4}
\tau_\mathrm{scat}\simeq\frac{6}{\pi a}\left(\frac{m_pcl_{c0}^2B_0^2}{2e}\right)^{1/3}\frac{1}{B}\left(\frac{\rho}{p}\right)^{1/3},
\end{equation}
where $m_p$ and $e$ are the proton mass and charge, respectively.

Where as the solar wind plasma is for the most part collision free, owing to its low density and high temperature, Coulomb collisions do play an important role in many interstellar environments. The timescale for momentum transfer via proton-proton collisions can be found in textbooks \citep[e.g.,][]{Burgers_1969}; in the present notations it is
\begin{equation}
\label{eq_tau_coll}
\tau_\mathrm{coll}=\frac{3m_p^2v^3}{8\sqrt{2\pi}ne^4\Lambda},
\end{equation}
where $\Lambda\sim 20$ is the Coulomb logarithm. Collisions in a bi-Maxwellian plasma lead to a decay of temperature anisotropy. In principle, the timescale depends on the anisotropy, but the dependence is sufficiently weak \citep{Hellinger_Travnícek_2010}, and our use of an isotropic expression for $\tau_\mathrm{coll}$ is justified.

Both pitch-angle scattering and proton-proton collision timescales depend on the environment; Table \ref{table_tau} lists their typical values at 30 au in the solar wind, at 100 au in the IHS, and at 130 au in the OHS for $a=0.2$.
\begin{table}[h]
\begin{center}
\caption{Characteristic pitch-angle scattering and proton-proton Coulomb collision timescales in the solar wind, the inner heliosheath, and the OHS.}
\label{table_tau}
\begin{tabular}{cccccccccc}
\tableline
             & $n$, cm$^{-3}$    &   T, K         & $B$, $\mu$G & $\Omega$, s$^{-1}$ & $v$, km s$^{-1}$ & $r_g$, au           & $l_c$, au & $\tau_\mathrm{scat}$, yr & $\tau_\mathrm{coll}$, yr \\
\tableline
SW (30 au)   & $6\times 10^{-3}$ &    $10^4$      &     1.5     &       0.014        &      13          & $6\times 10^{-6}$   &      0.2  &         0.02             &         4.4              \\
IHS (90 au)  &         $10^{-3}$ &    $10^5$      &     1       &       0.01         &      40          & $2.8\times 10^{-5}$ &      0.2  &         0.012            &       840                \\
OHS (130 au) &      0.1          & $2\times 10^4$ &     5       &       0.05         &      18          & $2.5\times 10^{-6}$ &   2000    &        5.4               &         0.75             \\
\tableline
\end{tabular}
\end{center}
\end{table}
Turbulent correlation lengths in the solar wind have been measured and modeled \citep{Breech_Matthaeus_Minnie_Bieber_Oughton_Smith_Isenberg_2008, Oughton_Matthaeus_Smith_Breech_Isenberg_2011}, and the value used in Table \ref{table_tau} is in agreement with the published values. To our knowledge, no such results for the IHS exist, so we use the solar-wind values as the nearest approximation. The correlation length for the OHS is based on the work of \citet{Burlaga_Florinski_Ness_2018}.

In the solar wind at 30 au, and to the lesser extent in the IHS, temperature equilibration is rapid because of strong wave-particle interactions; Coulomb collisions, by contrast, are infrequent and so should not affect the pressure anisotropy. In the OHS, however, collisional and scattering timescales are similar. In this work we set $\tau=\tau_\mathrm{scat}$ based on Table \ref{table_tau} in the entire heliosphere (solar wind and IHS). For the OHS and LISM, however, we explore a range of possible isotropization rates, including those from Table \ref{table_tau}, but also using larger and smaller values of this parameter (see the next section). To distinguish between the regions, a common approach is to use indicator variables $\alpha_i$ that are set to certain values at the boundary where the respective flow originates. In this model two indicator variables $\alpha_1$ and $\alpha_2$ are used that are initialized with a value of 1 at the inner and the outer boundary, respectively. The variables are passively advected with the plasma flow using the equation
\begin{equation}
\label{eq_alpha}
\frac{\partial(\rho\alpha_i)}{\partial t}+\nabla\cdot(\rho\alpha_i\mathbf{u})=0.
\end{equation}
One can then adopt a convention that values of $\alpha_i$ close to 1 (say $>0.9$) correspond to ``pure'' solar wind or pure LISM, while smaller values indicate that there is a possibility of both plasmas co-existing within a single numerical cell. More often than not such ``mixing'' is a consequence of a limited spatial resolution of the computational mesh. Most of the results on pressure anisotropy presented in this paper are restricted for the pure LISM region.

At present, the CGL model does not provide for coupling between protons and neutral hydrogen atoms via a charge exchange, which is an essential process in the solar wind-LISM interaction \citep{Blum_Fahr_1970, Holzer_1972, Wallis_Dryer_1976}. Charge exchange decreases the bulk solar wind speed, controls the geometric shape of the termination shock, and contributes to the bulk plasma pressure in both the IHS and the OHS \citep{Pauls_Zank_Williams_1995, Izmodenov_2000, Zank_Hunana_Mostafavi_Goldstein_2014}. To our knowledge, full energy and momentum charge exchange terms have not been calculated for a bi-Maxwellian ion distribution. The reader is therefore cautioned that the dimensions of the heliosphere and the shapes of the termination shock and heliopause in our single-fluid model might not agree with the models featuring neutral atoms using multi-fluid or fluid-kinetic descriptions \citep{Alexashov_Izmodenov_2005, Heerikhuisen_Florinski_Zank_2006}.


\section{Computational mesh, initial and boundary conditions} \label{sec:icbc}

As described in \citet{Bhoriya_Balsara_Florinski_Kumar_2024}, the governing system of equations is solved using second-order WENO reconstruction on a quadrilateral geodesic mesh, also known as a cube sphere \citep{Ronchi_Iacono_Paolucci_1996}. This mesh avoids the polar singularities of the spherical mesh and allows for a larger and more spatially uniform time step. Because of an absence of charge exchange, the inner boundary can be taken farther from the sun; we set the inner boundary at $r_\mathrm{min}=30$ au and the outer boundary at $r_\mathrm{max}=1000$ au. In the radial direction, the computational mesh consisted of 720 shells of variable width. Given a uniformly varying (index) variable $\xi$ that takes values between 0 and 1, the radial distance was computed as
\begin{equation}
\label{eq_radial_spacing}
r(\xi)=r_\mathrm{min}\left\lbrace1+\left[\frac{r_\mathrm{max}}{r_\mathrm{min}}-1-C(e^\alpha-1)\right]\xi+C(e^{\alpha\xi}-1)\right\rbrace,
\end{equation}
with the parameter
\begin{equation}
C=\left[\frac{r_\mathrm{med}}{r_\mathrm{min}}-1-\chi\left(\frac{r_\mathrm{max}}{r_\mathrm{min}}-1\right)\right][e^{\alpha\chi}-1-\chi(e^\alpha-1)]^{-1}.
\end{equation}
One can infer that the mapping given by (\ref{eq_radial_spacing}) results in an approximately uniform mesh between $r_\mathrm{min}$ and $r_\mathrm{med}$ followed by an exponentially rationed mesh at larger distances. The parameter $\chi$ gives the percentage of grid points that lie between $r_\mathrm{min}$ and $r_\mathrm{med}$, and $\alpha$ specifies how smooth the transition between the two regions is. In this paper we used $r_\mathrm{med}=250$ au, $\chi=0.4$ (40\% of grid points within the uniform shell region), and $\alpha=10$. The mesh on the sphere consisted of 29,400 quadrilateral faces.

A common practice in the global heliospheric modeling community is to start a simulation from a configuration consisting of two concentric annular regions. In this approach the inner region between $r_\mathrm{min}$ and $r_\mathrm{tr}$ (where ``tr'' stands for ``transition'') corresponds to the solar wind, and the outer region represents the LISM. The solar wind is described as a steady supersonic radial outflow, while the LISM is modeled as a uniform flow. The $z$ axis of the coordinate system used by the code is conventionally chosen to be parallel to the solar rotation axis $\hat{\mbox{\boldmath$\omega$}}$. The angle between the former direction and the LISM flow vector is close to $95^\circ$, which is often approximated as a right angle. The inflow direction is then chosen to be the $x$ axis. For the transition distance we took $r_\mathrm{tr}=150$ au.

The conditions in the solar wind are nominally specified at $r_1=1$ au and propagated to the inner boundary of the simulation using the assumption that the flow behaves adiabatically. The latter is, of course incorrect, owing to the production of the pickup ions by charge exchange that significantly increases the bulk temperature (and reduces the speed by some 10\%) of the combined plasma flow. To account for this, a temperature far in excess of the actual temperature at 1 au must be specified. In this work we used $2\times 10^7$ K at 1 au, which translates to a little over $2\times 10^5$ K at 30 au, which is not too far off from the New Horizons observations \citep{McComas_Zirnstein_Bzowski_Elliott_Randol_Schwadron_Sokol_Szalay_Olkin_Spencer_etal_2017}. The reference 1 au number density was 5 cm$^{-3}$, and the solar wind velocity was taken to be 450 km s$^{-2}$; these numbers were imposed at all latitudes meaning there was no distinction between the fast slow solar wind. The magnetic field was assumed to have a strength of $B_1=5$ nT and a winding angle of $45^\circ$ at 1 au in the solar equatorial plane; these values were propagated to $r_\mathrm{min}$ using the standard Parker solution \citep{Parker_1958}. The heliospheric current sheet was not included in the simulation. This is justified because the real current sheet oscillates in latitude producing a pattern of magnetic sectors that is too fine to be resolved, especially in the heliosheath \citep{Opher_Drake_Velli_Decker_Toth_2012, Pogorelov_Suess_Borovikov_Ebert_McComas_Zank_2013}. In addition, an inevitable numerical dissipation of magnetic field in the current sheet could lead to unphysically large values of the temperature anisotropy.

The heliospheric magnetic field field consists of two parts. The first is the radial monopole field spanning both the inner and the outer regions to ensure it is divergence-free. The use of a unipolar solar field in global heliospheric simulations is somewhat controversial \citep[e.g.,][]{Opher_Drake_Zieger_Gombosi_2015, Izmodenov_Alexashov_2015, Pogorelov_Borovikov_Heerikhuisen_Zhang_2015}, but the nature of the debate is largely irrelevant to the main thrust of this paper, and our choice gives us better control over the numerical experiment. Conventionally, to ensure that the initial magnetic field, which is defined at the faces of the computational cells, is divergence-free, it is initialized through the vector potential $\mathbf{A}$ on the edges of the respective face. The monopole field cannot be initialized this way because its vector potential has a singularity. Therefore, the radial field component
\begin{equation}
\label{eq_brad}
\mathbf{B}_\mathrm{rad}=\pm\frac{B_1 r_1^2}{\sqrt{2}}\frac{\mathbf{r}}{r^3}.
\end{equation}
Either polarity in equation (\ref{eq_brad}) can be chosen; the resulting model solution is independent of the sign in the absence of any field dissipation near the HP.

The second component is the azimuthal ﬁeld, restricted to the inner region. This field is given by
\begin{equation}
\label{eq_bint}
\mathbf{B}_\mathrm{azm}=\mp\frac{B_1 r_1}{\sqrt{2}}\frac{(\hat{\mbox{\boldmath$\omega$}}\times\mathbf{r})}{r^2},
\end{equation}
where the sign again depends on the chosen magnetic polarity in the unipolar model. The corresponding vector potential is
\begin{equation}
\mathbf{A}_\mathrm{azm}=\mp\frac{B_1 r_1}{\sqrt{2}}\left\lbrace\frac{\mathbf{r}}{r}+\frac{[(\hat{\mbox{\boldmath $\omega$}}\times\mathbf{r})\times\mathbf{r}]}{r^2}\right\rbrace.
\end{equation}
We remind the reader that in our notation $\hat{\mbox{\boldmath $\omega$}}=\hat\mathbf{e}_z$.

The region $r>r_\mathrm{tr}$ represents the environment of the low density warm interstellar cloud surrounding the solar system known as the Local Interstellar Cloud (LIC). The initial analytic solution in the exterior region is that of an incompressible, potential flow around a solid sphere. The velocity vector is tangential to the surface of the sphere at $r=r_\mathrm{tr}$ and tend to a uniform state at $r\to\infty$. Thus $\rho_\mathrm{ext}=\rho_\infty$, $p_\mathrm{ext}=p_\infty$, where the ``infinity'' subscript refers to pristine interstellar values, and
\begin{equation}
\mathbf{u}_\mathrm{ext}=\left(1+\frac{r_\mathrm{tr}^3}{2r^3}\right)\mathbf{u}_\infty-\frac{3r_\mathrm{tr}^3}{2r^5}(\mathbf{u}_\infty\cdot\mathbf{r})\mathbf{r},
\end{equation}
where $\mathbf{u}_\infty=-u_\infty\hat\mathbf{e}_x$ is the interstellar inflow vector. The interstellar magnetic field is several orders of magnitude stronger than the initial interior field at $r_\mathrm{tr}$, and the initial interaction between the two flows can create very strong discontinuities that could lead to unstable behavior of the CGL code, which is more sensitive to flow gradients that conventional MHD codes. For this reason we start with a tapered initial magnetic field
\begin{equation}
\label{eq_bext}
\mathbf{B}_\mathrm{ext}=\frac{B_\infty}{2}(1+\tanh\psi)\hat\mathbf{b},
\end{equation}
where $\hat\mathbf{b}$ is the chosen direction of the magnetic field at infinity, and
\begin{equation}
\label{eq_psi}
\psi=\frac{\hat\mathbf{n}\cdot(\mathbf{r}-\mathbf{u}_\infty t)-r_{1/2}}{\delta}.
\end{equation}
In equation (\ref{eq_psi}) $\hat\mathbf{n}$ is a unit vector normal to the planes $B=\mathrm{const}$, $r_{1/2}$ is the distance from the origin to the plane where the field's strength decreases to one half of its interstellar value, $t$ is time, and $\delta$ is the taper width. We used $r_{1/2}=225$ au and $\delta=15$ au. The corresponding exterior vector potential is
\begin{equation}
\label{eq_aext}
\mathbf{A}_\mathrm{ext}=\frac{B_\infty\delta}{2}[\psi+\log(\cosh\psi)](\hat\mathbf{b}\times\hat\mathbf{n}).
\end{equation}
The numerical values of the pristine LISM parameters were: $B_\infty=3\;\mu$G, $n_\infty=0.1$ cm$^{-3}$, $T_\infty=7\times 10^3$ K, and $u_\infty=26$ km s$^{-1}$. The magnetic field made an angle of
$140^\circ$ with the LISM flow direction and the BV plane (the plane containing the LISM flow velocity and magnetic field vectors) was tilted by $45^\circ$ relative to the meridional plane. These values are consistent with our current knowledge of the LISM properties \citep{Zirnstein_Heerikhuisen_Funsten_Livadiotis_McComas_Pogorelov_2016}.

At the inner boundary the uniform and steady solar-wind conditions were maintained in the ghost zones throughout the simulation. In the outer boundary's ghost zones the expressions for the magnetic field and vector potential were updated in time according to equations (\ref{eq_bext}) and (\ref{eq_aext}) to account for the advection of the initial profile. In the tail region the so-called non-reflecting boundary conditions were used to accelerate the flow to the local fast magnetosonic speed \citep{Pogorelov_Semenov_1997}. To simplify the problem we used the isotropic version of the boundary conditions based on the averaged pressure $p=(p_\parallel+2p_\perp)/3$.


\section{Heliospheric structure: MHD vs. CGL} \label{sec:results}
A total of three simulations were performed using different values of the pressure relaxation parameter in the LISM region. In the first simulation the timescale $\tau_\mathrm{scat}$ was 100 times smaller than estimated in Table \ref{eq_tau_coll}. This corresponds, essentially, to the isotropic MHD case. The second simulation used the nominal values, and the third used a relaxation time 10 times longer than nominal. The last case would correspond to a situation where the turbulence in the OHS was unusually weak or almost entirely two-dimensional (and therefore ineffective in scattering the particles), while the temperature in the outer heliosheath was higher by a factor of about 5, so that the mean collision time was increased tenfold. Note that the LISM was recently shown to be highly inhomogeneous \citep{Swaczyna_Schwadron_Mobius_Bzowski_Frisch_Linsky_McComas_Rahmanifard_Redfield_Winslow_etal_2022}. It is therefore likely that the heliosphere was exposed to very different interstellar conditions in the past. All simulations were run to at least 400 years, which should be sufficient to establish a nearly steady flow pattern.

\begin{figure}[h]
\begin{center}
\includegraphics[width=0.42\textwidth,clip=]{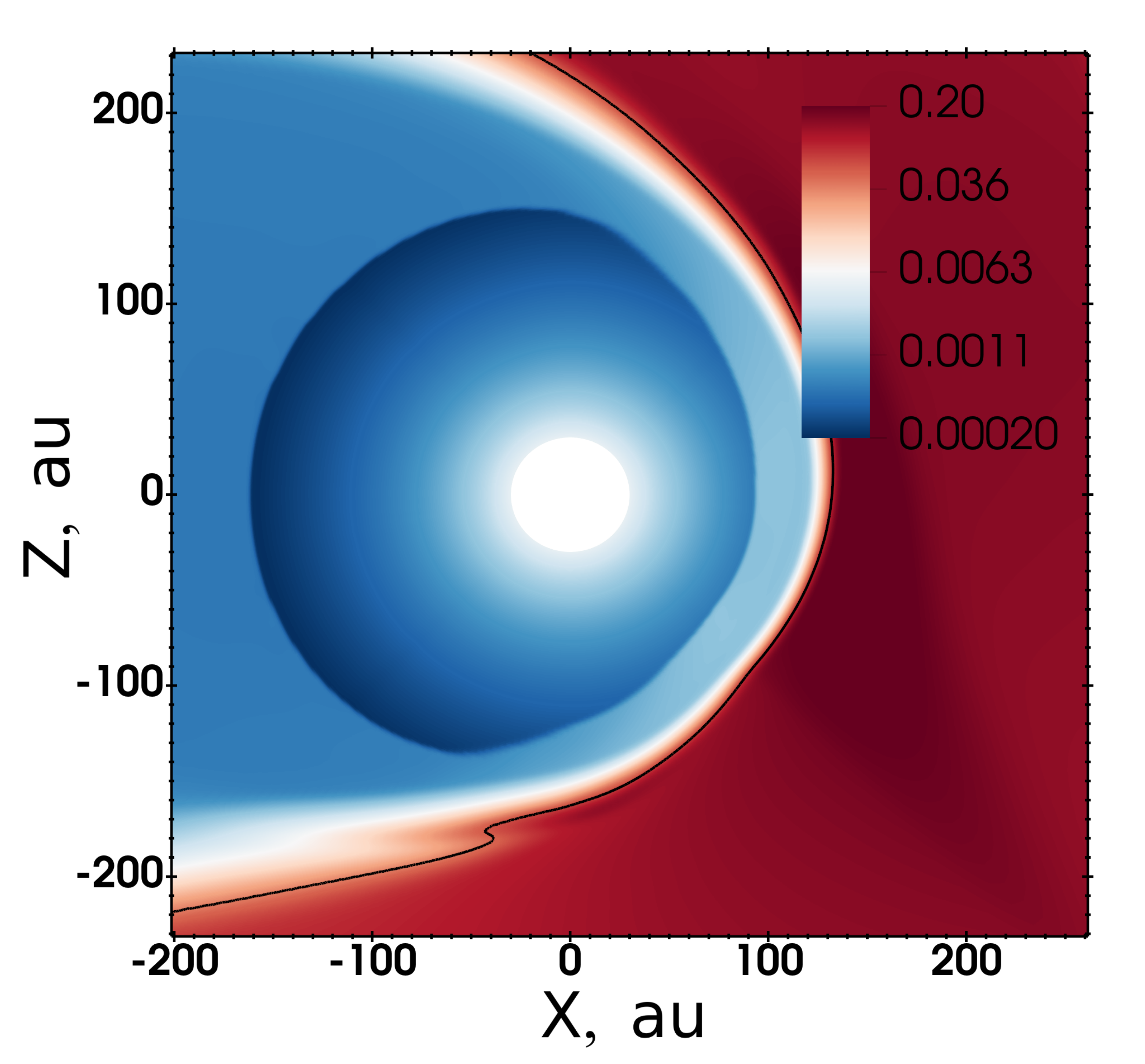}
\includegraphics[width=0.42\textwidth,clip=]{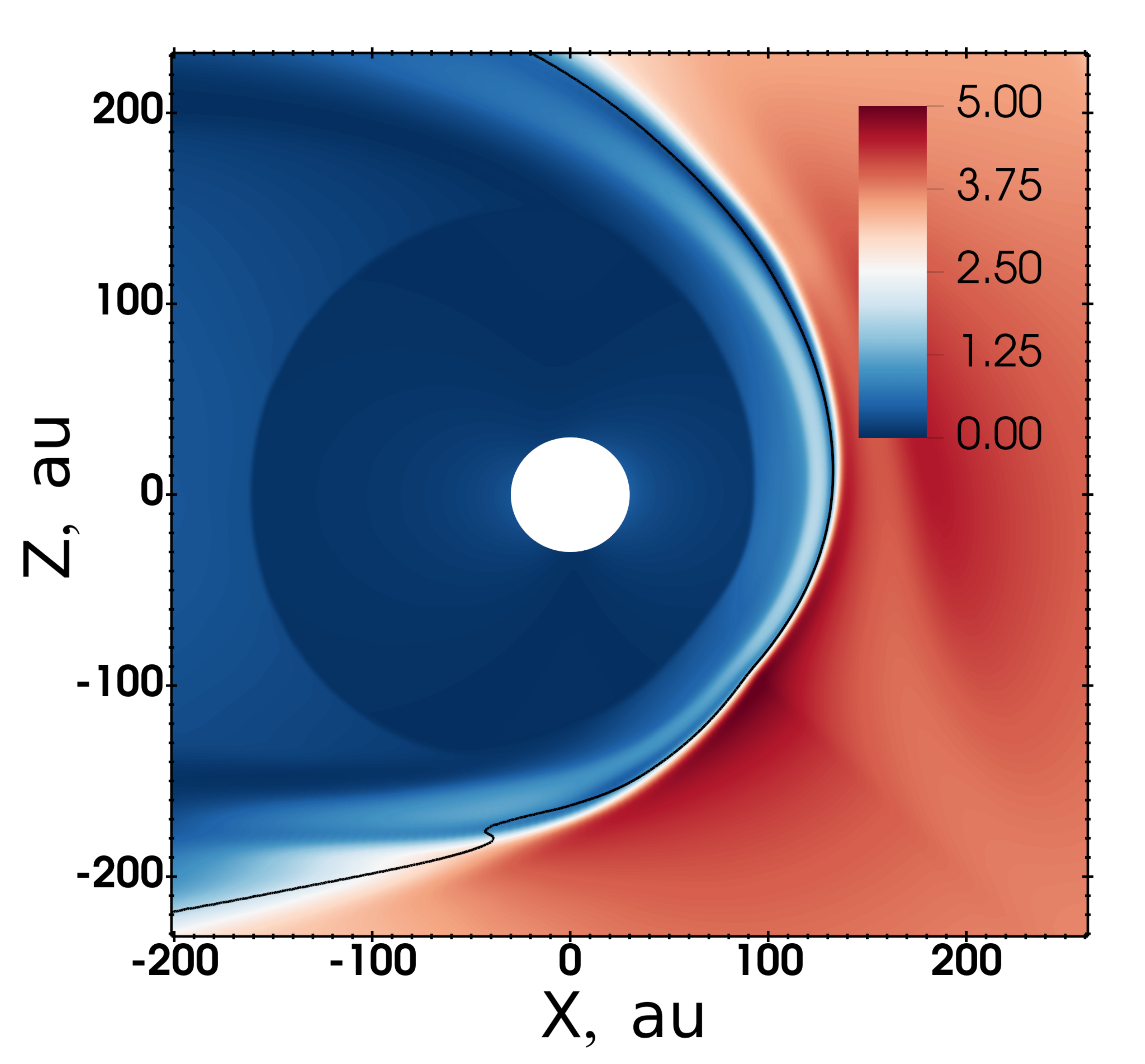}
\includegraphics[width=0.42\textwidth,clip=]{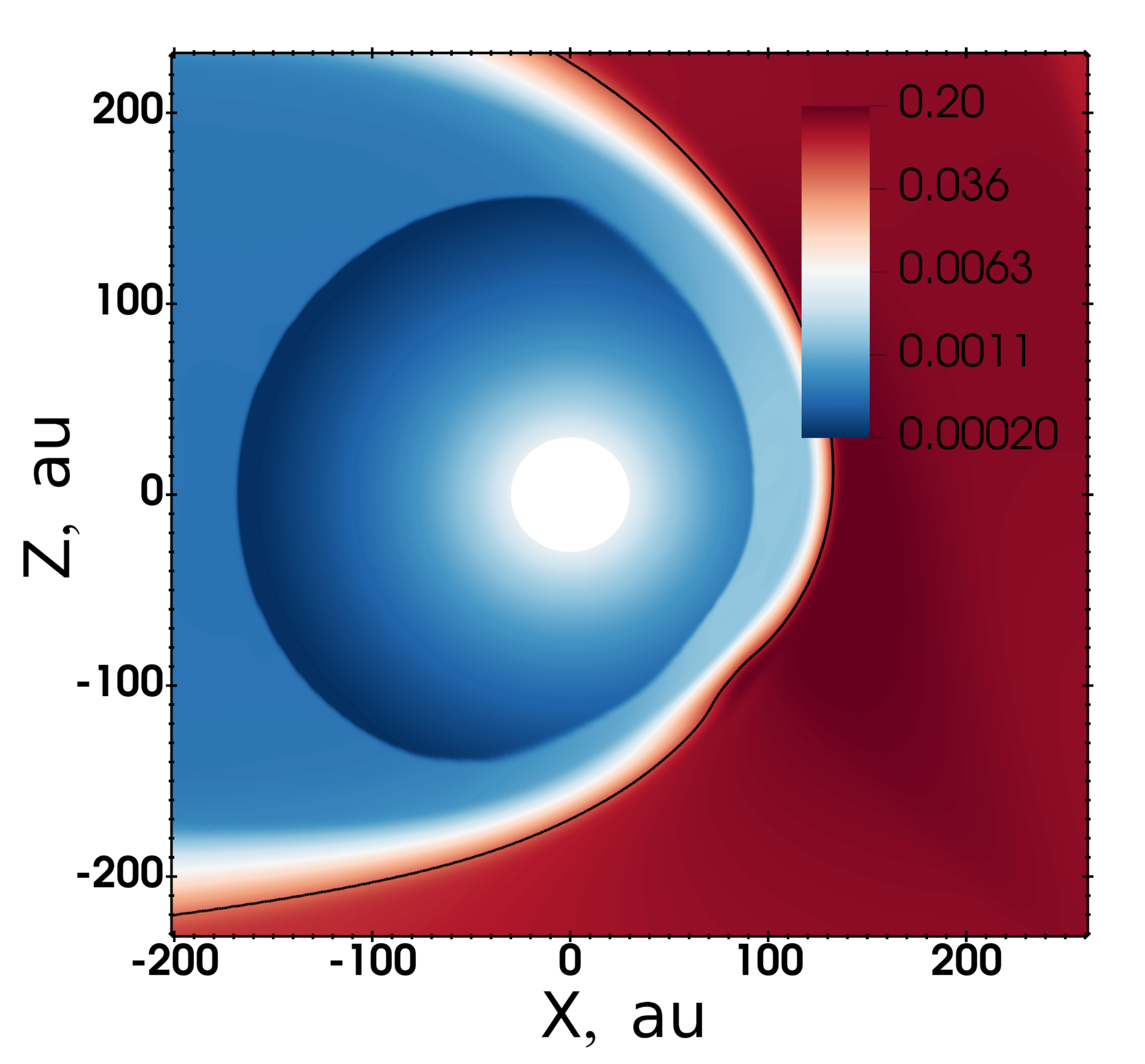}
\includegraphics[width=0.42\textwidth,clip=]{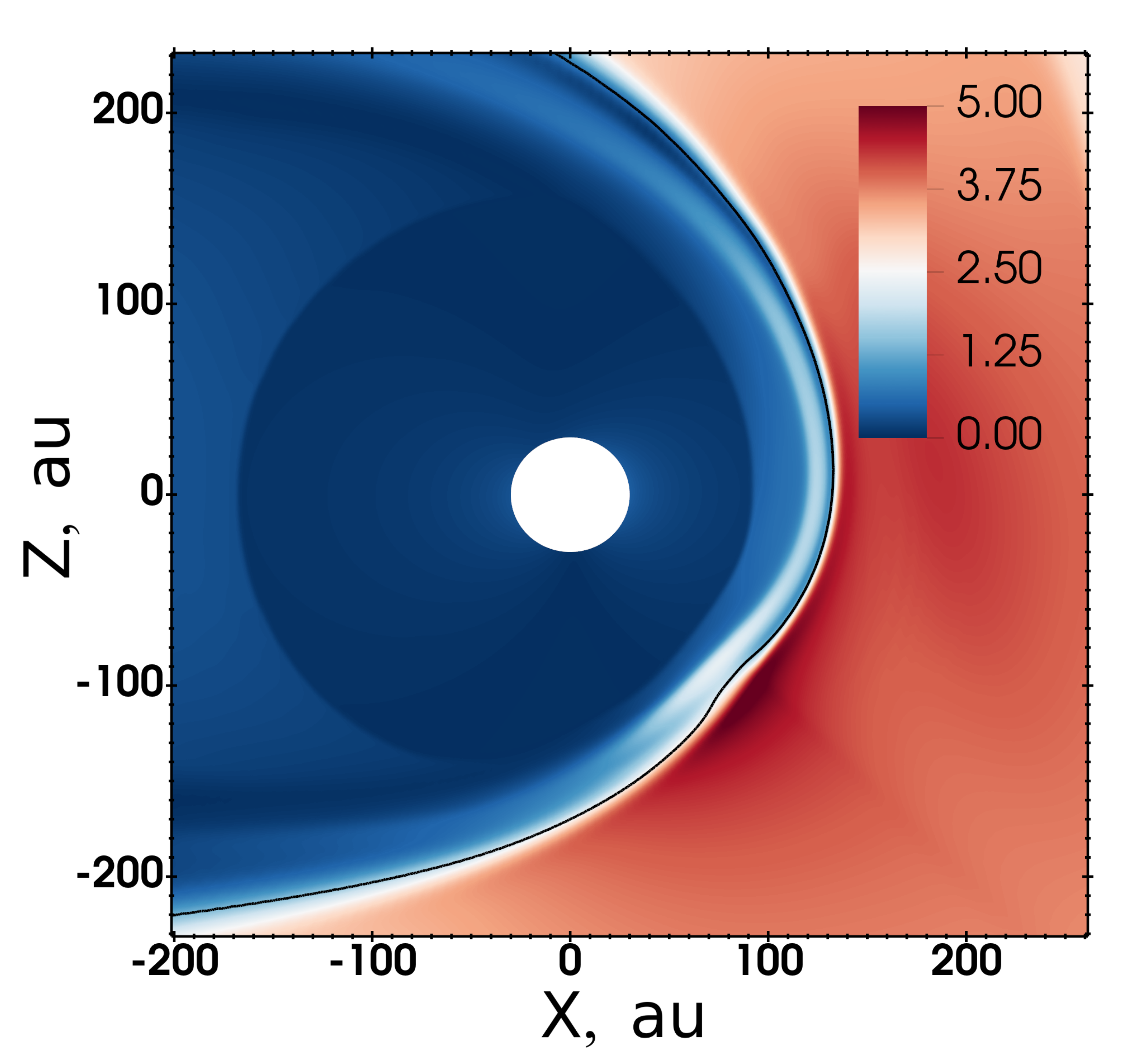}
\includegraphics[width=0.42\textwidth,clip=]{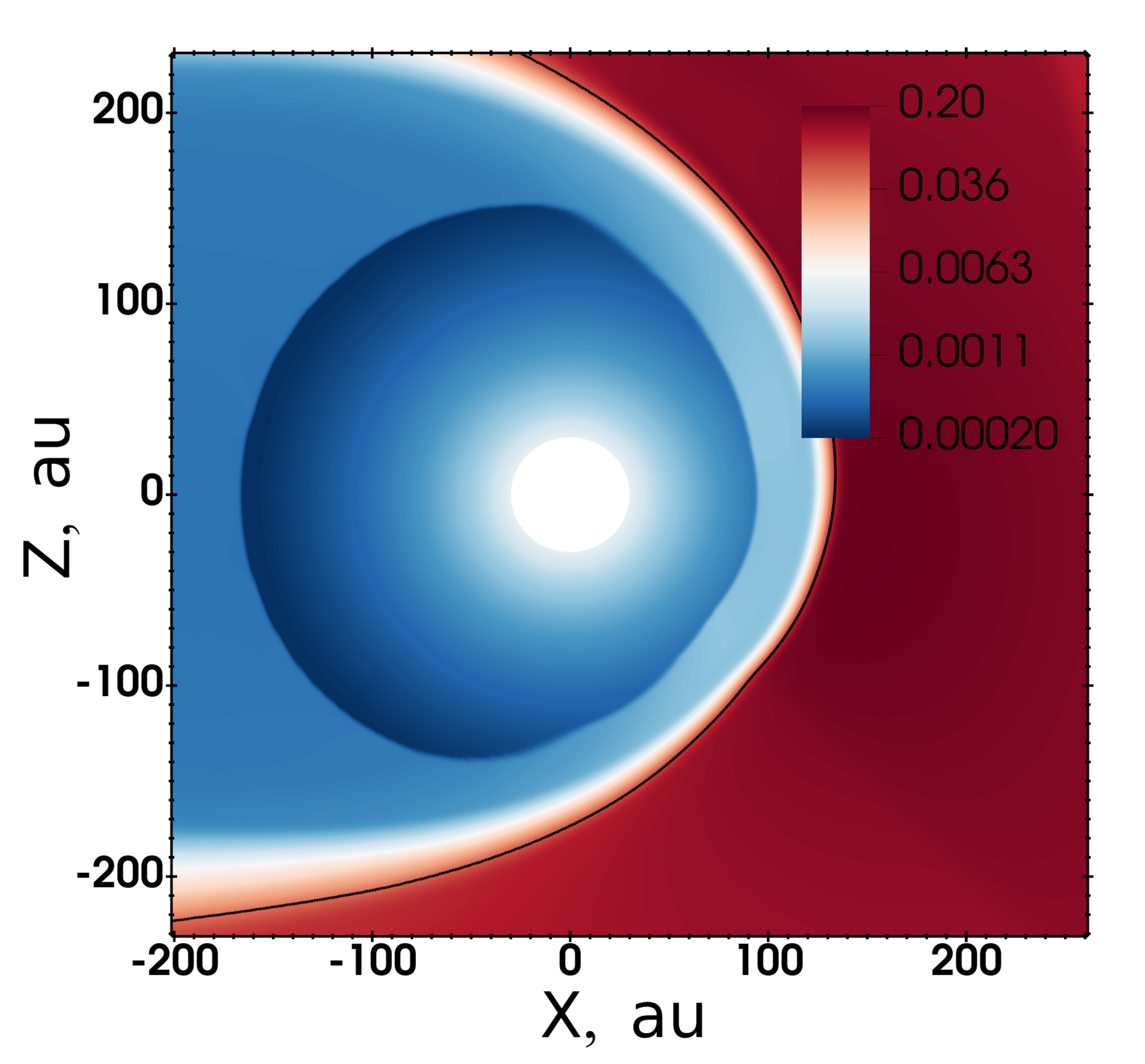}
\includegraphics[width=0.42\textwidth,clip=]{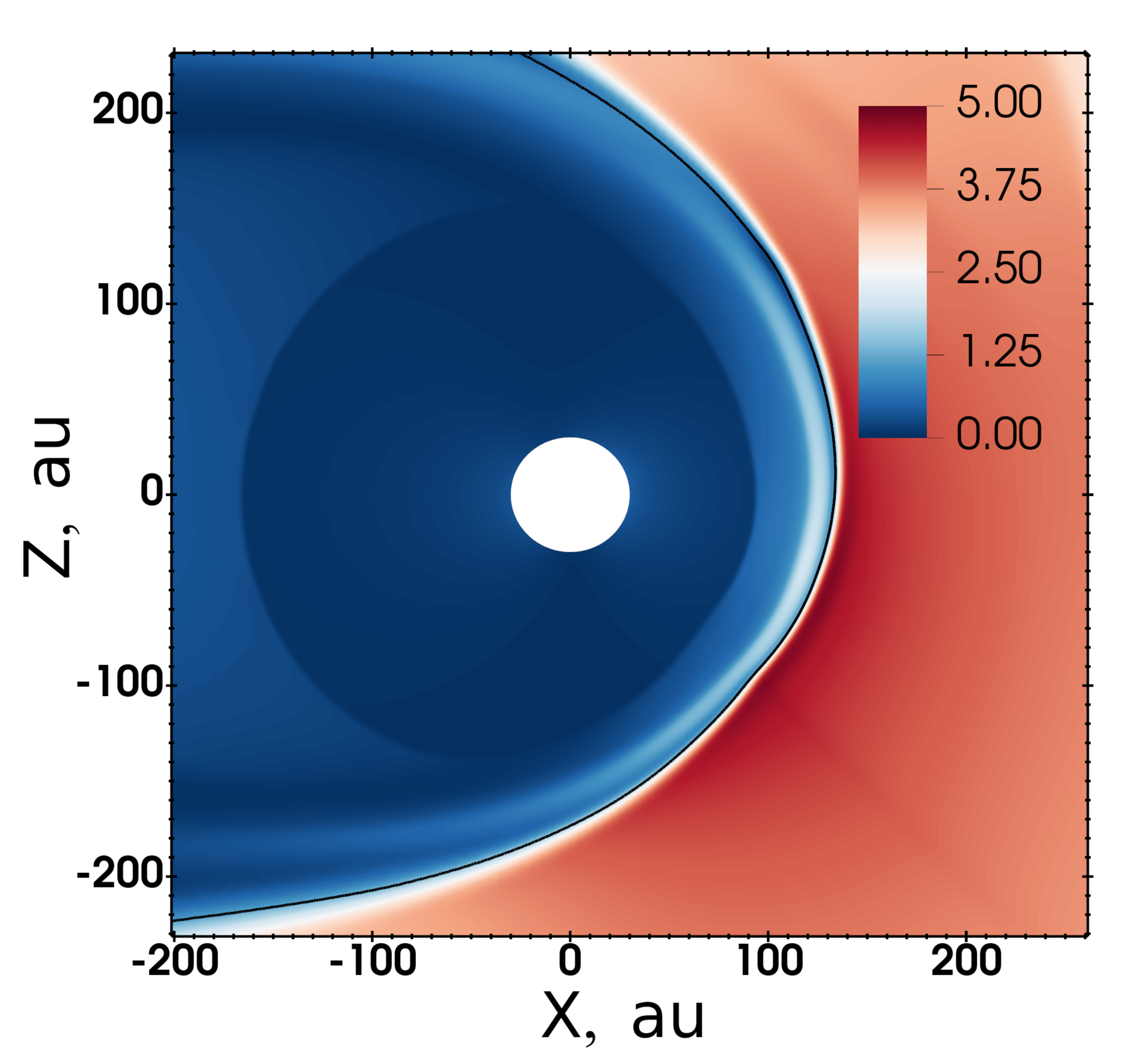}
\caption{Plasma density $n$ in cm$^{-3}$ (left panels) and magnetic field strength $B$ in $\mu$G (right panels) in the meridional ($xz$) plane for the quasi-MHD (top), nominal (middle) and low (bottom) scattering rate. The figure was produced with VisIt software \citep{HPV_VisIt}. The black line is the contour of $\alpha_2=0.99$, which is considered the edge of the heliosphere.}
\label{fig_den_mag}
\end{center}
\end{figure}

Figure \ref{fig_den_mag} shows the plasma number density in the left panels and the magnetic field strength $B$ in the right panels, in the meridional plane, which is the plane containing the solar rotation axis and the LISM flow vector. In the density plots the red color corresponds to the LISM, while the less dense solar wind region is blue in color. The solid black line is the contour $\alpha_2=0.99$, which we treat as the definition of the heliopause. Note that only the forward (``nose'') portion of the heliosphere is shown in the figure; most of the heliotail is not covered by a high-resolution mesh so the differences between the models are not as dramatic.

The termination shock (easily discernible in the plots as a rapid change in color inside the heliospheric region) has a prominent nose-tail asymmetry that is characteristic of models without neutral atoms. Note, however, that both the termination shock and HP are not fully stationary, but feature slowly evolving surface irregularities. For example, the small ``kink'' seen in the contours of the indicator variable is traveling slowly along the surface of the HP. This could be a result of an instability, but it remains to be seen whether it will disappear or strengthen once the hydrogen atoms are added to the system. The minor color change at about $45^\circ$ (toward the lower right corner) is a mesh imprint effect. It could be eliminated, if needed, by using spatial reconstruction of order higher than two, but the effect is not large enough to justify such an effort. The bow shock exists in this configuration with a compression ration of 1.3--1.4. The shock is expected to be even weaker with the inclusion of charge exchange as a consequence of loading of the OHS plasma by secondary pickup ions born from energetic neutrals in the IHS \citep{Zank_Heerikhuisen_Wood_Pogorelov_Zirnstein_McComas_2013}.

\begin{figure}[h]
\begin{center}
\includegraphics[width=0.42\textwidth,clip=]{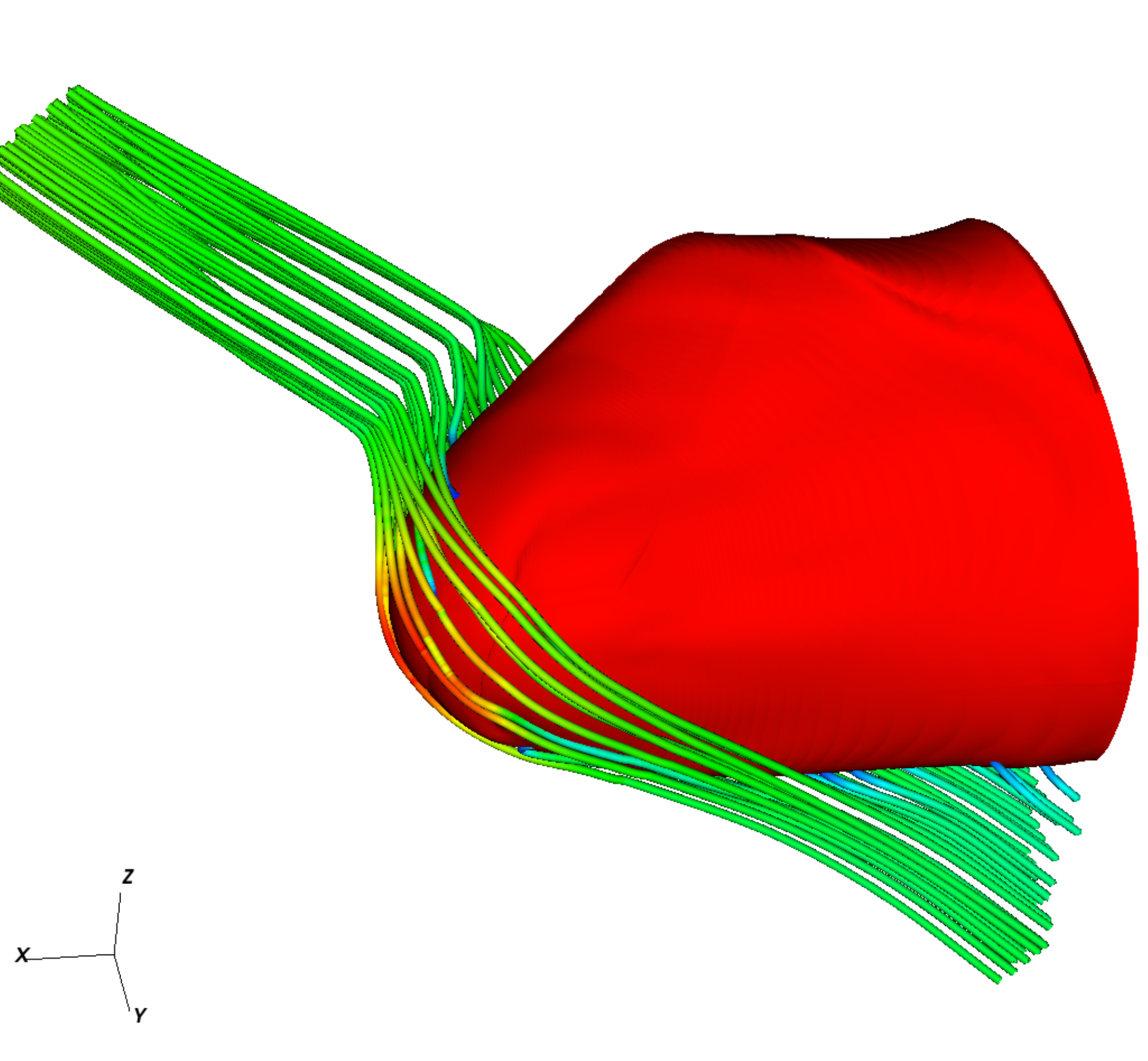}
\includegraphics[width=0.42\textwidth,clip=]{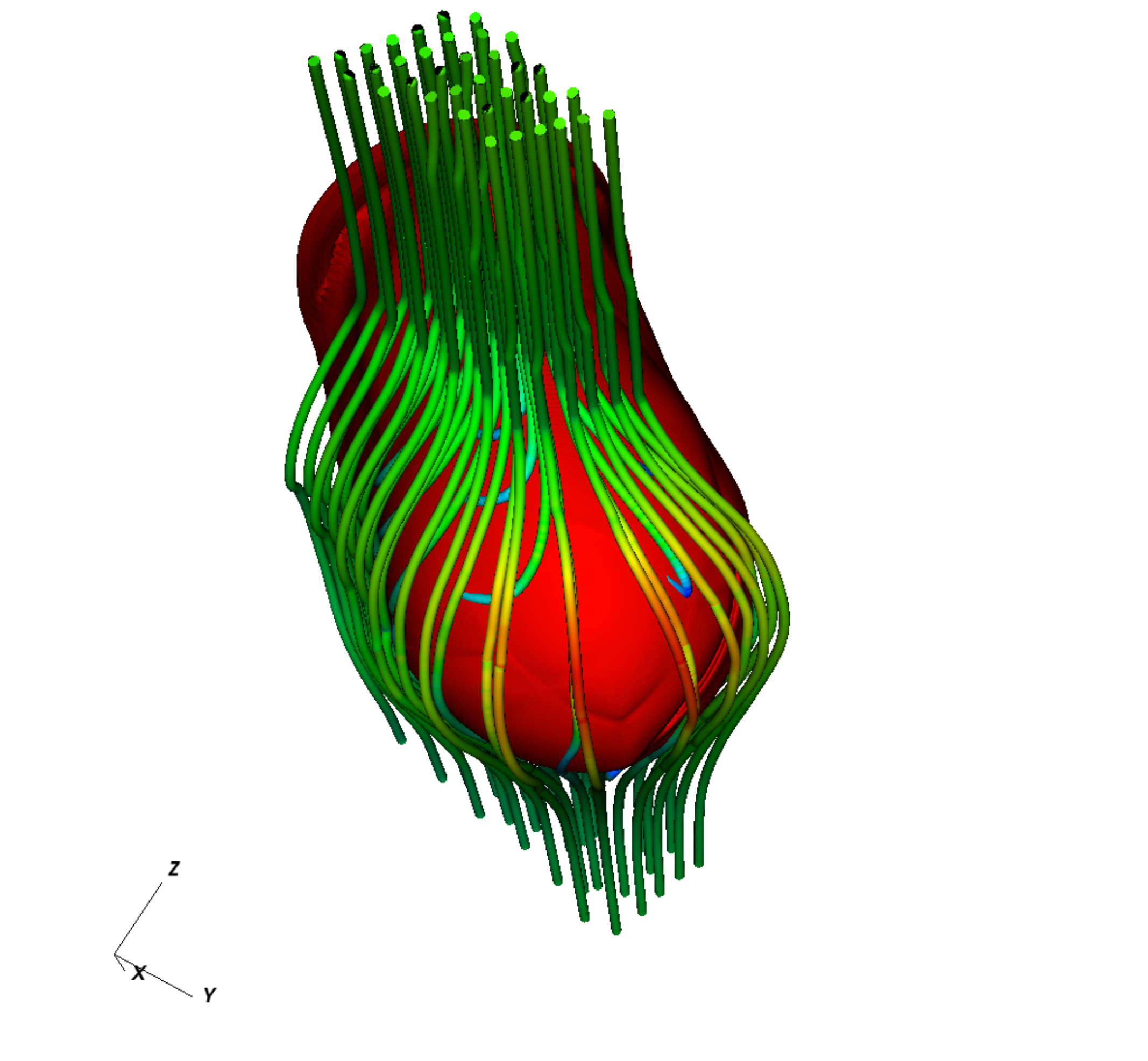}
\caption{Magnetic field draping pattern around the ``windward'' part of the heliopause computed from the low scattering rate simulation. The red structure is the surface of the HP. Individual magnetic fields are shown as flux tubes with the color proportional to the magnetic field strength.}
\label{fig_draping}
\end{center}
\end{figure}

Figure \ref{fig_draping} illustrates the draping of the magnetic field around the HP as viewed from the flank (left) and the front (right). The HP (shown in red) has a typical structure familiar from  MHD models \citep{Zirnstein_Heerikhuisen_Funsten_Livadiotis_McComas_Pogorelov_2016, Izmodenov_Alexashov_2020}. The heliosphere is compressed in the direction perpendicular to the LISM magnetic field, and the structure is rotated relative to the BV plane. As discussed in \citet{Florinski_AlonsoGuzman_Kleimann_Baliukin_Ghanbari_Turner_Zieger_Kota_Opher_Izmodenov_etal_2024}, the magnetic field is strongest in the south where the flow pushes it against the HP, and weakest in the north, where the flow is more parallel to the field lines. As show below, these anisotropy should behave differently in these regions, $A>0$ in the south, and $A<0$ in the north.

\begin{figure}[h]
\begin{center}
\includegraphics[width=0.42\textwidth,clip=]{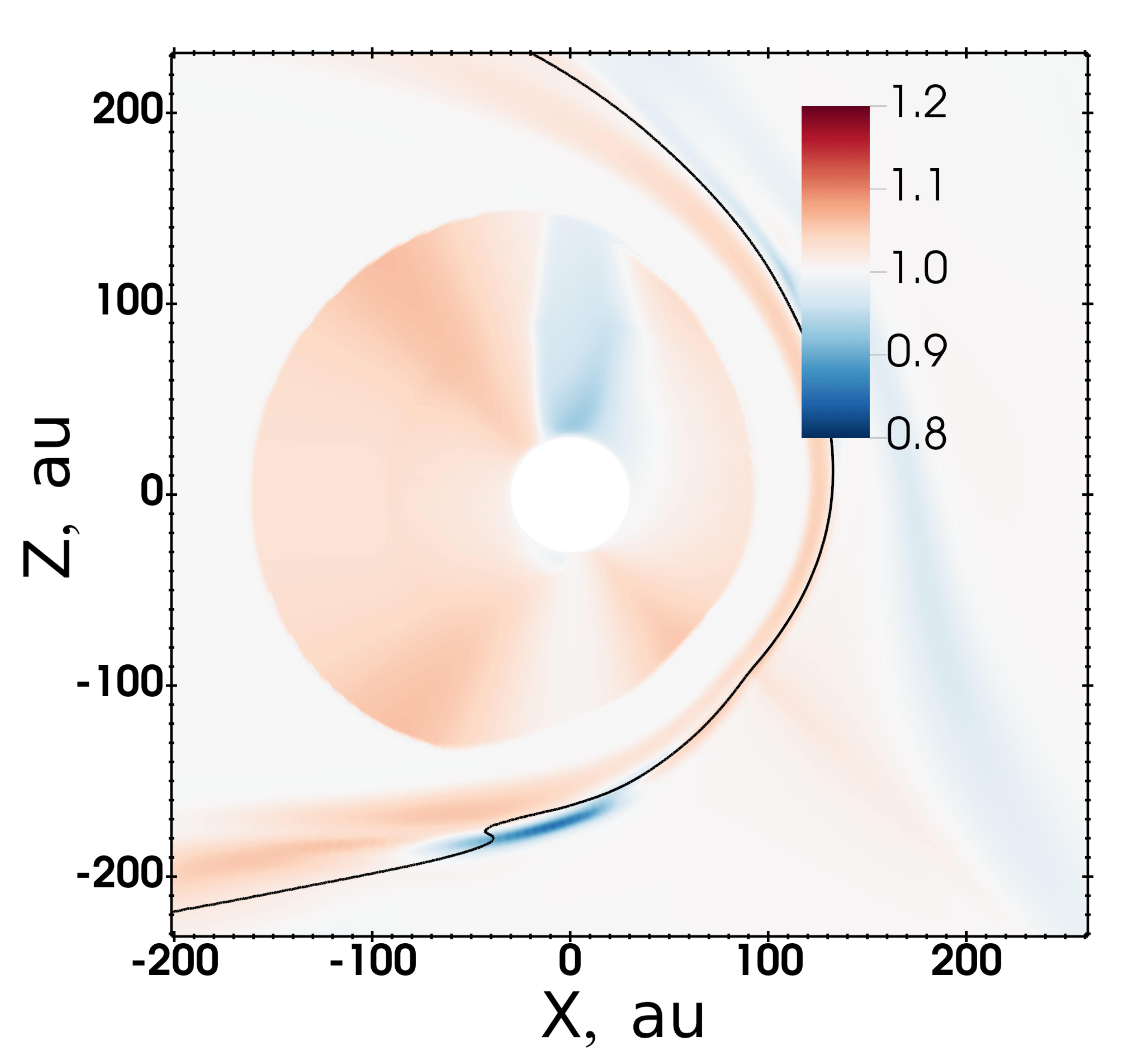}
\includegraphics[width=0.42\textwidth,clip=]{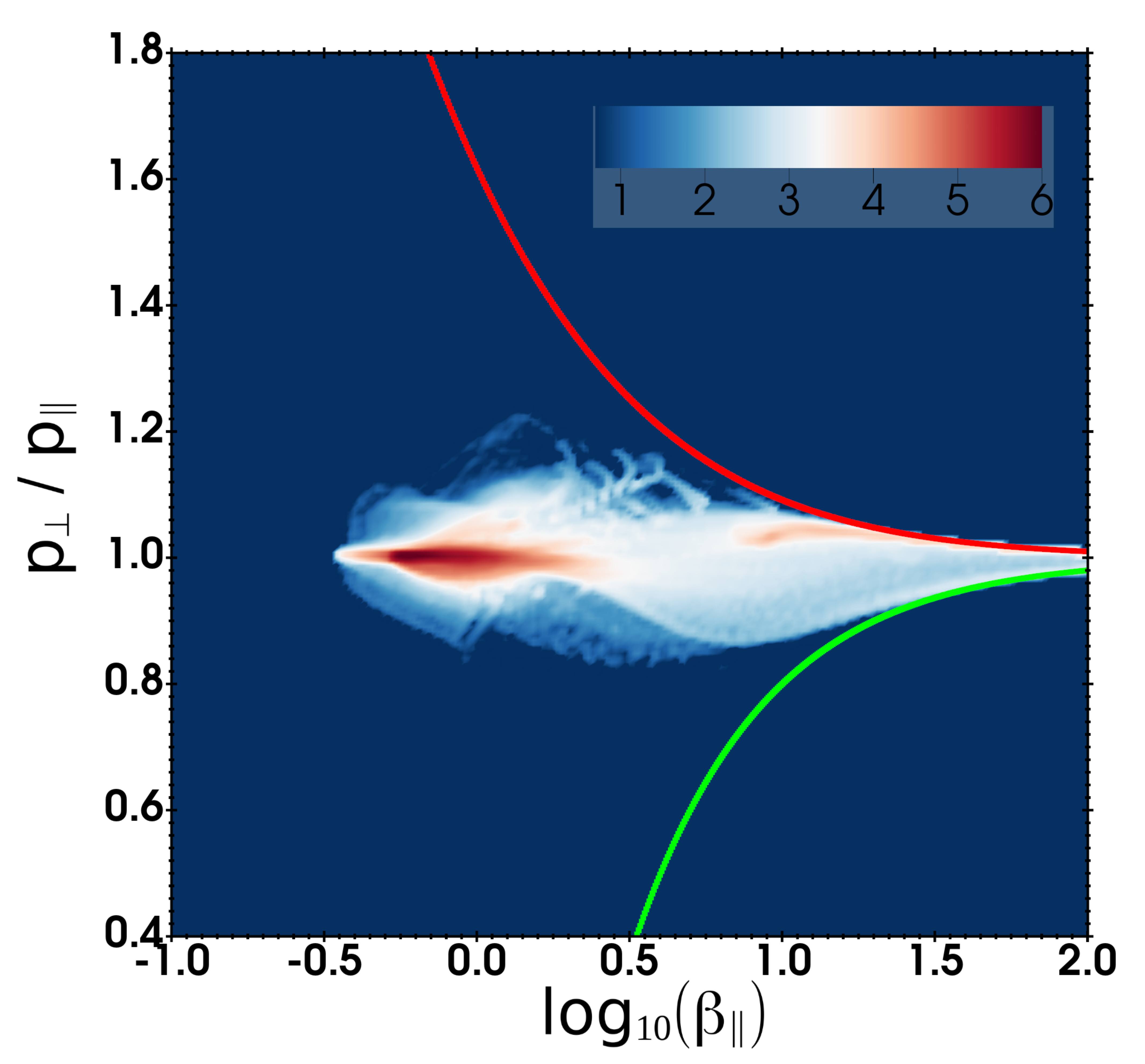}
\includegraphics[width=0.42\textwidth,clip=]{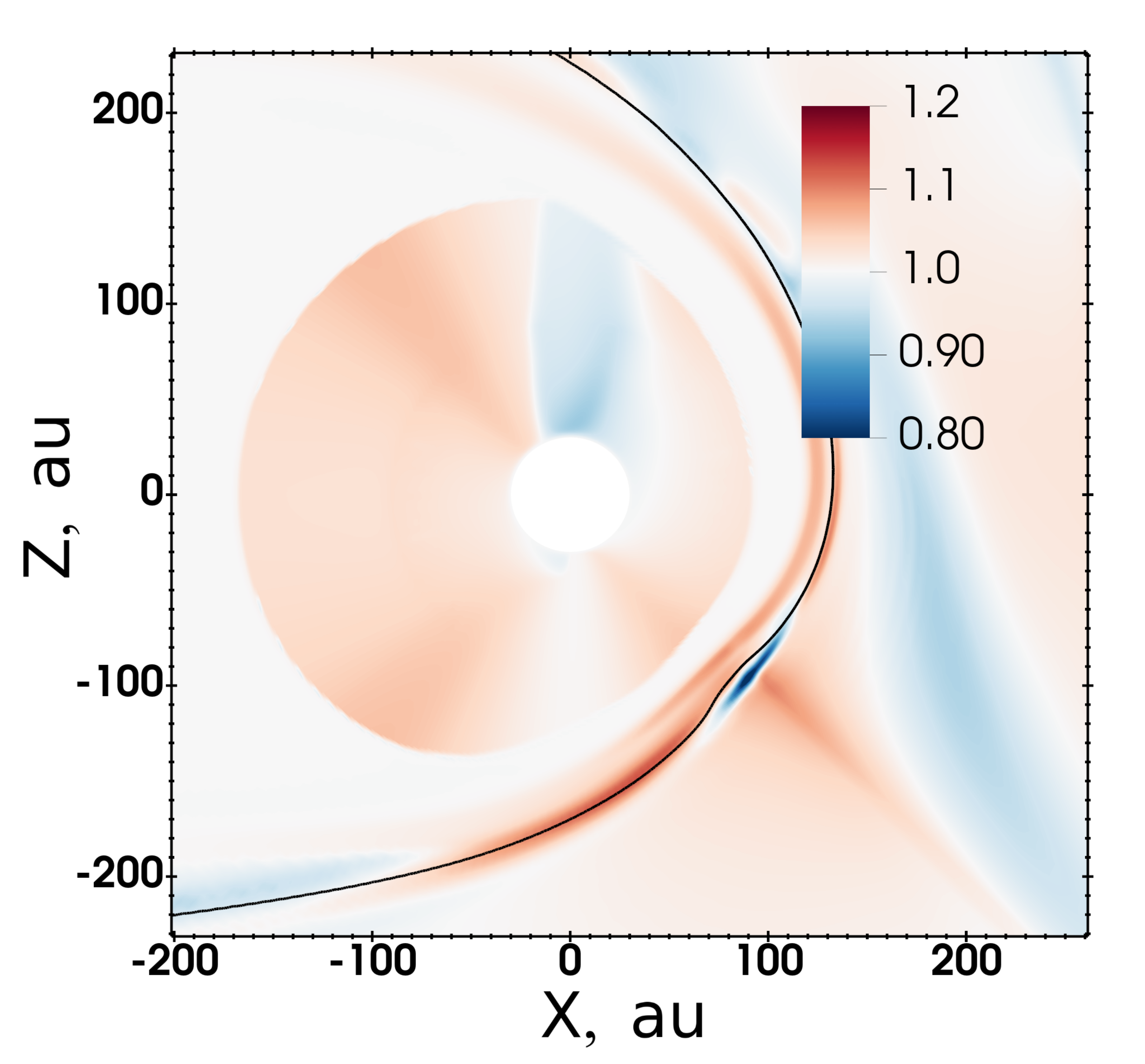}
\includegraphics[width=0.42\textwidth,clip=]{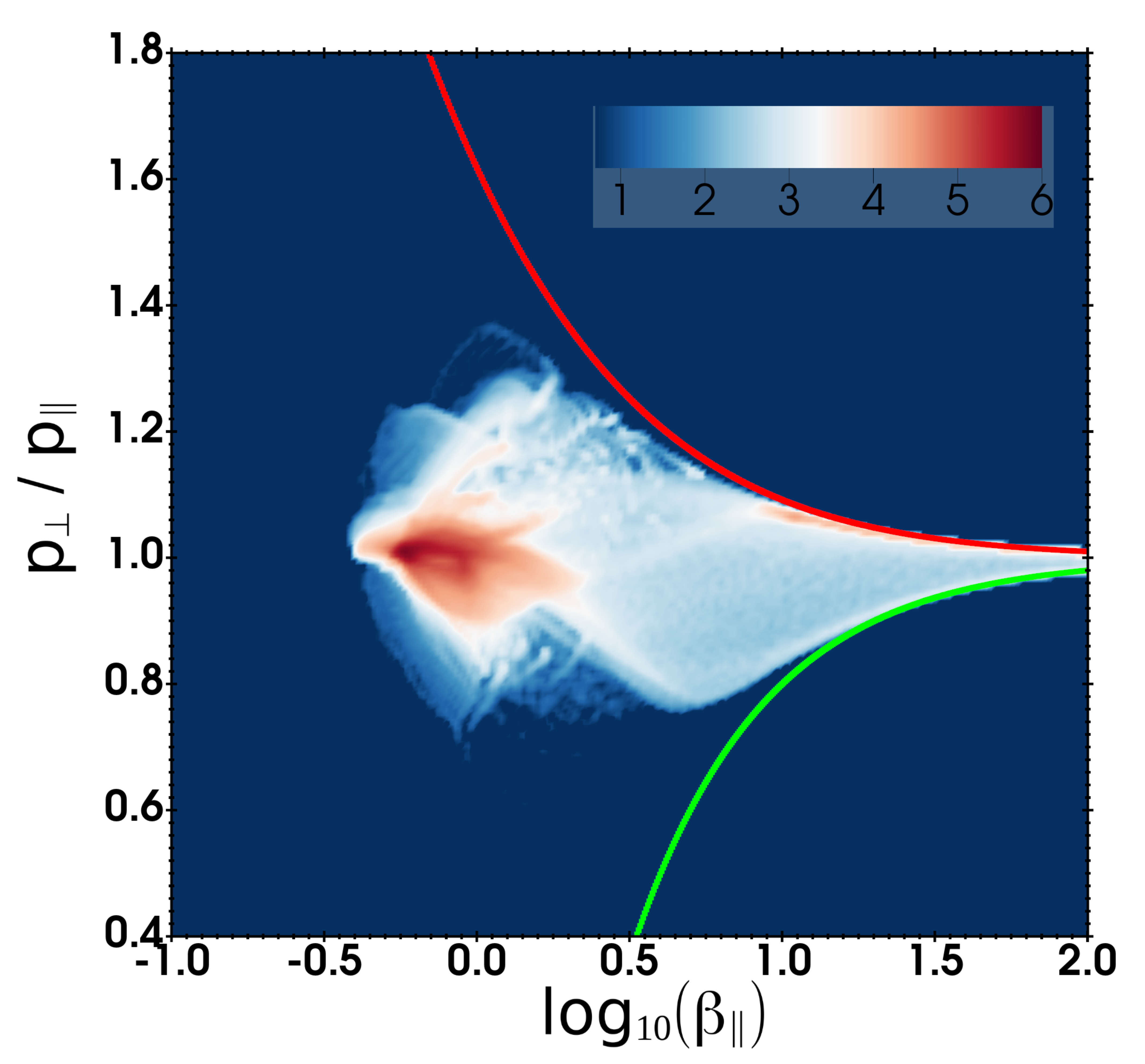}
\includegraphics[width=0.42\textwidth,clip=]{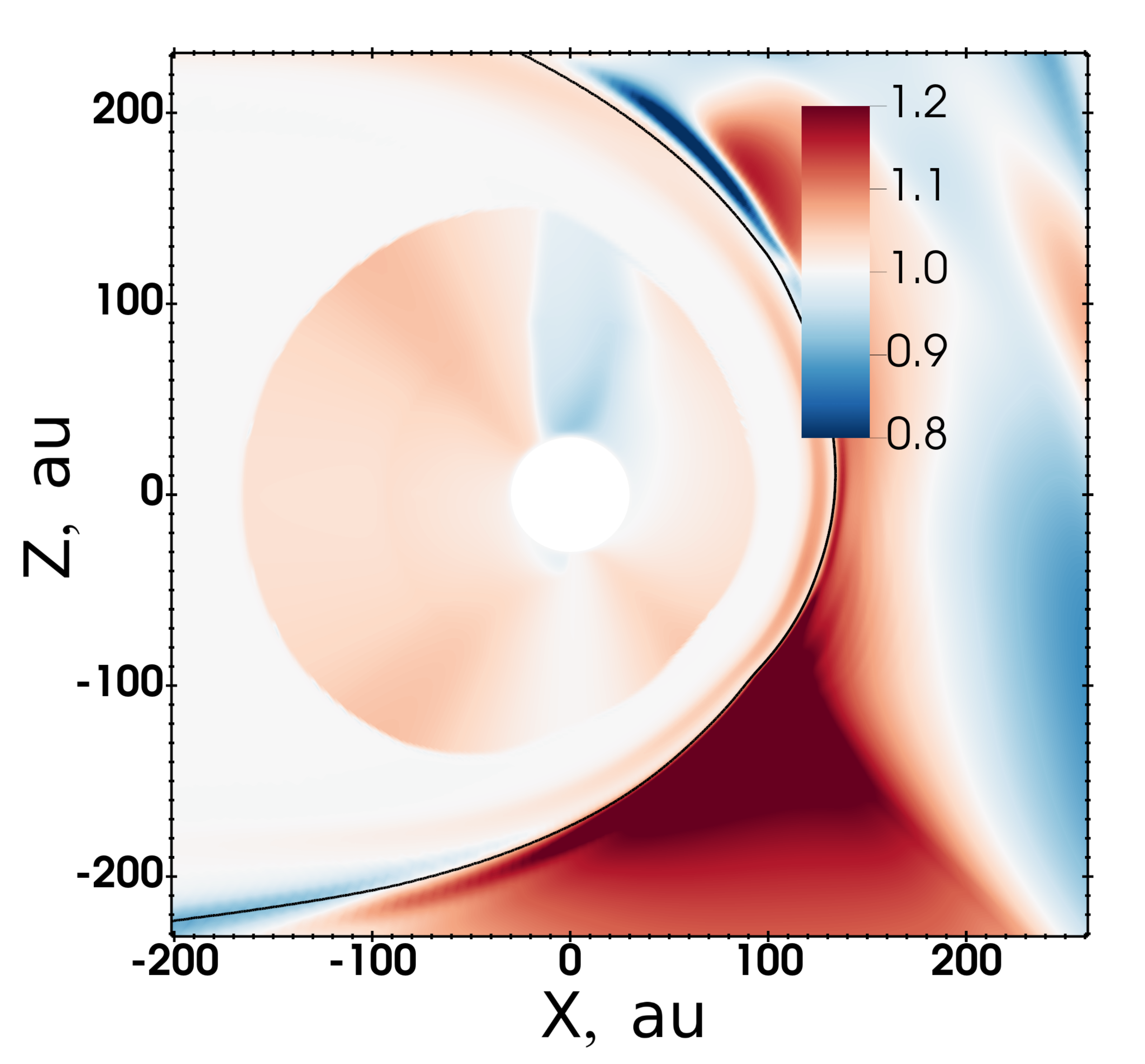}
\includegraphics[width=0.42\textwidth,clip=]{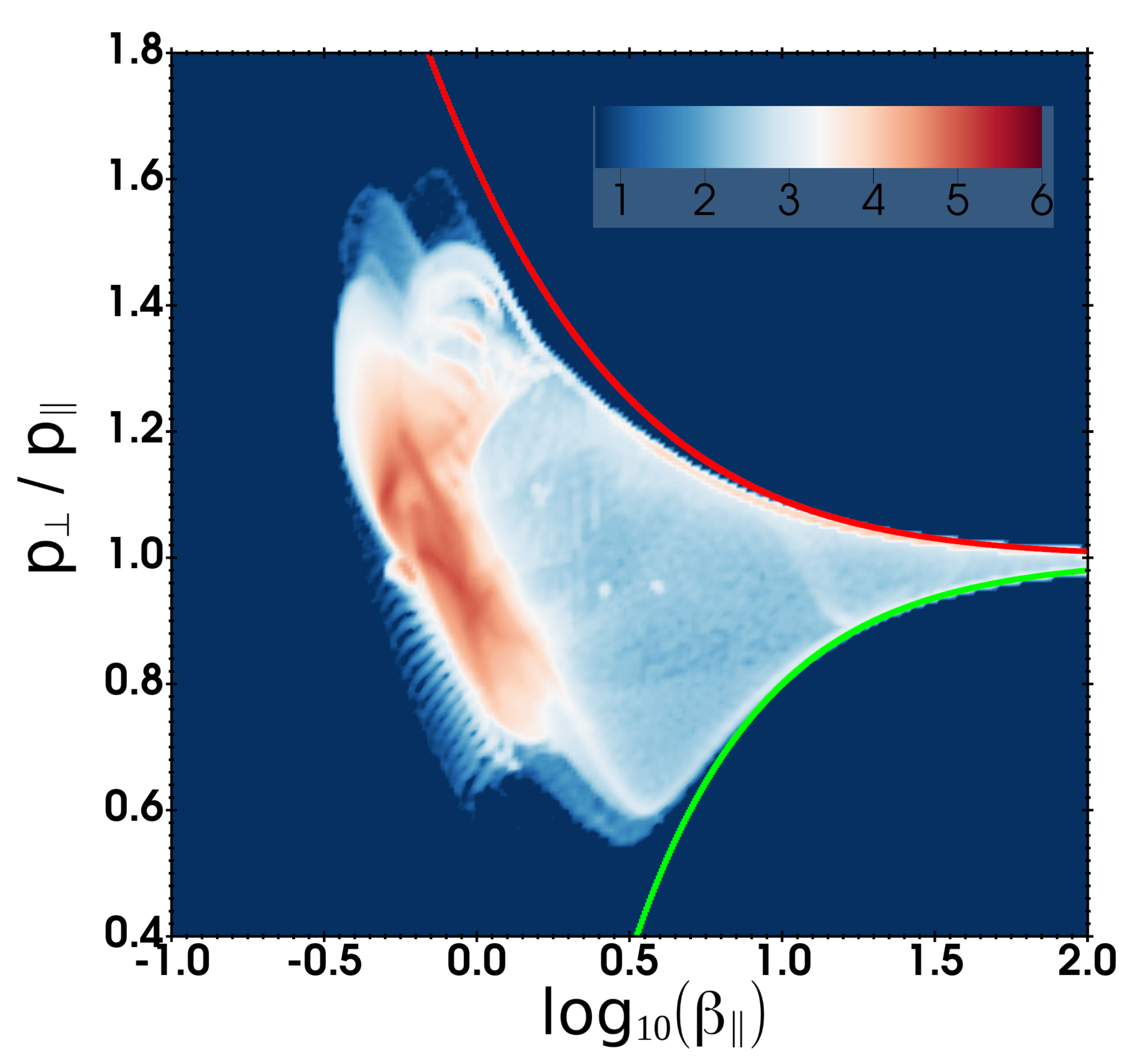}
\caption{Pressure anisotropy $A$ in the meridional plane (left panels) and the binned distribution of the pressure anisotropy values vs. parallel plasma beta (right panels) for the quasi-MHD (top), nominal (middle) and low (bottom) scattering rate. The binned distributions were generated for the nose region defined by $\alpha_2>0.9$, $r<300$ au, $x>0$. The $A(\beta_\parallel)$ distributions were weighted by the cell volume and use logarithmic color scale. The green lines refer to the firehose instability threshold given by equation (\ref{eq_threshold_firehose}) and the red lines refer to the mirror instability threshold, equation (\ref{eq_threshold_mirror}).}
\label{fig_aniz}
\end{center}
\end{figure}

Figure \ref{fig_aniz} shows our results for the pressure anisotropy. The left panels show the expected increasing trend of $A$ with increased isotropization time from top to bottom. The MHD-like simulation (upper panel) is essentially isotropic, while the ``standard'' case has a narrow region of $A<1$ in the north. This result can be understood as follows. The magnetic field topology in that region is cusp-like, and the plasma flows mainly parallel to the field lines leading to a density increase; because the deformation is along the direction of $\mathbf{B}$, it leads to an increase in parallel pressure. The low-scattering simulation features a broad region with $A>0$ in the south. Here the plasma compression occurs primarily normal to the magnetic field lines, and an excess in $p_\perp$ develops. This feature bears a direct similarity with magnetospheric PDLs. The thickness of the simulated anisotropic layer is between 5 and 10 astronomical units, which is roughly in line with theoretical estimates \citep{Fraternale_Pogorelov_2021}. The last (extreme) case features a massive anisotropic layer extending hundreds of au into the LISM. The anisotropy in the layer varies between $A=0.7$ in the magnetic cusp region in the north and $A=1.6$ in the magnetic compression region in the south.

The frequency distribution of pressure anisotropy in the solar wind is conventionally presented on a $A(\beta_\parallel)$ diagram \citep{Bale_Kasper_Howes_Quataert_Salem_Sundkvist_2009, Hellinger_Travnicek_Kasper_Lazarus_2006}. In a similar spirit, we present distributions of \textit{spatial locations} with a given pressure ratio as a function of the parallel plasma beta. The right panels of Figure \ref{fig_aniz} shows the distributions for all three cases. Each was generated using $200\times 200$ bins using logarithmic bin spacing in $\beta_\parallel$ and linear spacing in the anisotropy. Because we are interested in the upwind OHS, and not in the IHS or the LISM region upstream of the bow shock (both of which are essentially isotropic), we restricted the binning algorithm to only count the cells that satisfy $\alpha_2>0.9$ (to exclude the heliosphere), $r<300$ au (to exclude the pre-bow-shock LISM) and, and $x>0$ (to exclude the tail region). The count rates were weighted by the cell volume. The dark red region corresponds to the ``core'' LISM plasma that is isotropic and has $\beta\sim 0.5$. As expected, the simulation with a hundredfold increase in the scattering rate has the anisotropy distribution that is strongly concentrated near the line $A=1$. The nominal and especially the low scattering rate simulations show much larger anisotropy variations in the OHS, ranging from $\sim 0.5$ and up to $1.6$. For the latter, the dark red ``core'' of the distribution is also spread between 0.8 and 1.2, which implies that most of the plasma in the upstream OHS exists in an asisotropic state. The red and green lines show the mirror and firehose instability thresholds, respectively (see equations \ref{eq_threshold_firehose} and \ref{eq_threshold_mirror}); as expected the numerical results are contained to the left of these curves.


\section{Discussion and implications for extrasolar PDLs} \label{sec:discuss}
Every numerical study of the interaction between the solar system and its plasma environment to date was grounded in the assumption that the plasma distribution was isotropic in all space. This paper presents a major departure from this paradigm by allowing the interstellar plasma to develop a bi-Maxwellian distribution with distinctly different $p_\parallel$ and $p_\perp$. The anisotropy develops as a result of flow deformation in a directions parallel, perpendicular, or oblique to the magnetic field. A flow compression normal to the field leads to an increase in the perpendicular temperature as a result of conservation of magnetic moment. Conversely, a compression acting along the field lines tends to increase the parallel pressure owing to conservation of the longitudinal adiabatic invariant. Scattering on magnetic field fluctuations is the mechanism that prevents the plasma from developing prohibitively large anisotropies and crossing the stability threshold, which is forbidden in fluid models. The model used in this study demonstrated stable behavior under a very wide range of scattering rates.

Viewed on global scales, the interaction between the heliosphere and the LISM is not dramatically affected by the introduction of a pressure anisotropy. The flow and magnetic field draping patterns are quite similar in all three of our simulations, including the MHD-like. Plasma interactions with interstellar atoms and energetic particles (pickup ions and cosmic rays) could have a more pronounced effect on the structure \citep{Fahr_Kausch_Scherer_2000, Myasnikov_Alexashov_Izmodenov_Chalov_2000}. However, unlike MHD, the CGL model can tell us where pressure anisotropy driven instabilities are likely to occur and deduce the properties of the turbulence produced by these instabilities.

A common trend in the pressure anisotropy just beyond the HP is a development of a $A<1$ region in the northern hemisphere and a $A>1$ region in the south relatively to the solar equatorial plane. This behavior is fully consistent with the expectations based on the orientation of the magnetic field in the LISM and the direction of deformation, as discussed above. This implies that plasma in the northern hemisphere could develop a firehose instability leading to a generation of Alfven or magnetosonic waves. Conversely, the southern hemisphere features a structure similar to a planetary plasma depletion layer, with a dominant perpendicular pressure component. We expect that mirror mode or ion cyclotron waves will be generated if the mirror instability threshold is crossed by the plasma. It is possible that these differences could be confirmed experimentally by comparing the compressibility of the turbulent magnetic fluctuations at Voyager 1 and Voyager 2 that are traveling in the northern and the southern hemispheres, respectively.

The lack of charge exchange in the model prevents us from making a quantitative comparison with the Voyager observations. For example, the LISM plasma density is the single fluid model is necessarily higher than measured; as a result the OHS plasma has a density of $\sim 0.2$ cm$^{-3}$, which is well in excess of the observed values. Charge exchange can, in principle, be treated with the isotropic model using the average pressure $p=(p_\parallel+2p_\perp)/3$. Even in that case equation (\ref{eq_deltap}) would have to be modified to incorporate the effects of replacing protons taken from a bi-Maxwellian distribution with Maxwellian-distributed pickup ions. In our view, however, a framework built on an anisotropic plasma description should employ properly formulated anisotropic physics of charge exchange. This requires substantial additional theoretical development, and we leave it to future work.

While this work was focused on the OHS, there is some evidence that the IHS plasma is also anisotropic. \citet{Liu_Richardson_Belcher_Kasper_2007} argued that the plasma downstream of a quasi-perpendicular termination shock has $A>1$, and should be susceptible to a mirror instability. \citet{Tsurutani_Lakhina_Verkhoglyadova_Echer_Guarnieri_Narita_Constantinescu_2011} and \citet{Burlaga_Ness_2011} interpreted the pressure-balanced structures (exhibiting anticorrelation between magnetic field strength and plasma density) observed by Voyager 1 in the IHS as mirror mode structures. Additionally, \citet{Fichtner_Kleimann_Yoon_Scherer_Oughton_Engelbrecht_2020} performed a calculation of anisotropy evolution based on quasi-linear interaction with the instability-generated mirror mode fluctuations in the IHS, starting from initially imposed values for $A$. The advantage of a CGL model is that the pressure anisotropy is provided as a part of the solution. However, we only include ambient magnetic fluctuations and ignore the possible instability-generated waves. Another physical aspect that is missing from the present model is the pickup process that tends to preferentially increase the perpendicular pressure of the bulk plasma in the IHS (in the nose direction), where the magnetic field is mainly perpendicular to the flow of interstellar neutral hydrogen. One can therefore envisage a two-fluid CGL model, where the pickup ions and the core thermal protons both have bi-Maxwellian distributions with different values of $A$.

Based on the simulation results from the model with a long isotropization timescale, a much broader PDL (hundreds of au, see the bottom left panel of Figure \ref{fig_aniz}) could exist around astrospheres of other stars provided the conditions in the outer astrosheath result in somewhat weaker wave-article interactions, but also in a tenfold increase in the mean time between Coulomb collisions. The former can be readily achieved if the turbulence outer scale were larger by a factor of two (compared with Table \ref{table_tau}) for the same magnetic field strength. The outer scale is thought to be related to the size of the astrosphere \citep{Burlaga_Florinski_Ness_2018}, which presumably can vary by orders of magnitude depending on the rate of mass loss from the star and the density of the surrounding LISM. The proton-proton collision rate could be reduced by a factor of ten if the bow shock were strong enough to heat the interstellar plasma to temperatures in excess of $3\times 10^4$ K. This, in turn, would require that the star had a sufficiently large velocity relative to its circumstellar environment. An elementary estimate (ignoring the effects of the magnetic field) shows that a speed about twice the local sound speed in the LISM would be sufficient to heat the medium to the required temperature. A larger compression ratio would increase the density downstream of the bow shock, and to compensate for this the LISM density would have to be smaller by a factor of 2 compared with the density of the LIC.

\citet{Muller_Frisch_Florinski_Zank_2006} modeled several astrospherical configurations whose parameter range satisfy the above constraints. Their models 20--25 explored the effects of changes in the cloud-star relative velocity for diffuse, partially ionized interstellar clouds similar to the LIC. Most of those cases, however, correspond to small astrospheres, so the conditions on the turbulence outer scale might not hold there. \citet{Scherer_Fichtner_Heber_Ferreira_Potgieter_2008} considered a scenario when a solar-like star was embedded in a tenuous interstellar cloud (their case ``b''). That would be a good candidate for an astrospheric configuration with a thick depletion layer. Another interesting situation where collisional effects would be depressed is the environment of a low density and hot interstellar ``bubbles'' evacuated by supernova explosions such as the Local Bubble \citep{Frisch_2007}. In that case the bow shock does not exist, but the temperature in the bubble is already hot enough (up to $10^6$ K) to make Coulomb collisions totally ineffective. The PDL in that case would be governed by wave-particle interactions and be strongly dependent on turbulent processes at the astrospheric interface.

\begin{acknowledgments}
VF was supported by NASA grants 80NSSC20K0786 and by NSF grant 2009871. DSB was supported by NSF grants 2009776, 2434532, NASA grant NASA-2020-1241, and NASA grant 80NSSC22K0628. Computer simulations were performed at the Center for Research Computing cluster at UND, and at the Regional Computing Hub for Alabama Universities at UAH funded by NSF grant 2232873.
\end{acknowledgments}

\bibliography{manuscript}{}
\bibliographystyle{aasjournal}

\end{document}